\documentclass[aps,pre,floats,twocolumn,showpacs]{revtex4-1}
\usepackage{graphicx}
\usepackage{graphics,dcolumn,bm,fleqn,epic,eepic,float}
\usepackage{amssymb,amsmath,multirow,rotate,color}
\usepackage{subfigure}
\usepackage{color}

\usepackage{amssymb,amsfonts,amsmath,multirow}
\usepackage{verbatim}

\newcommand{\be}{\begin{equation}}
\newcommand{\ee}{\end{equation}}
\newcommand{\bea}{\begin{eqnarray}}
\newcommand{\eea}{\end{eqnarray}}
\newcommand{\bit}{\begin{itemize}}
\newcommand{\eit}{\end{itemize}}

\renewcommand{\emph}{\textbf}

\begin{document}
\title{A tale of many cities: universal patterns in human urban mobility}


\author{Anastasios Noulas$^{1}$, Salvatore Scellato$^{1}$, Renaud
  Lambiotte$^{2}$, Massimiliano Pontil$^{3}$, Cecilia
  Mascolo$^{1}$}

\affiliation{%
  $^1$~University of Cambridge, Cambridge,UK}
\affiliation{%
  $^2$~University of Namur, Namur, Belgium}
\affiliation{%
  $^3$~University College London, London, UK}

\date{\today}
\begin{abstract} 
The advent of geographic online social networks such as Foursquare, where users
voluntarily signal their current location, opens the door
to powerful studies on human movement. In particular the fine granularity of
the location data, with GPS accuracy down to 10 meters, and the worldwide scale of
Foursquare adoption are unprecedented. In this paper we 
study urban mobility patterns of people in several metropolitan
cities around the globe by analyzing a large set of Foursquare users.
Surprisingly, while there are variations in human movement in different cities,
our analysis shows that those are predominantly due to heterogeneous distribution of places across different urban environments.  
Moreover, a universal law for human mobility is identified, which isolates the rank distance as a key component,
factoring in the number of places between origin and destination,
rather than pure physical distance, as considered in previous works. Building on
our findings, we also show that a rank-based movement accurately
captures real human movements in different cities. Our results shed new
light on the driving factors of urban human mobility, with potential
applications to urban planning, location-based advertisement and even
social studies.
\end{abstract}
\maketitle
\begin{figure*}[t]
\subfigure
{
\includegraphics[scale =1] {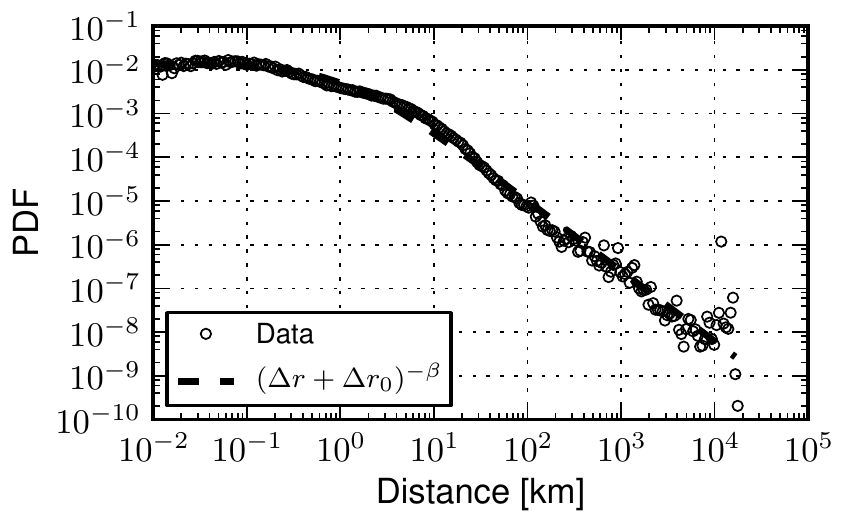}
\label{alltrans}
 }
\subfigure
{
\includegraphics[scale =1] {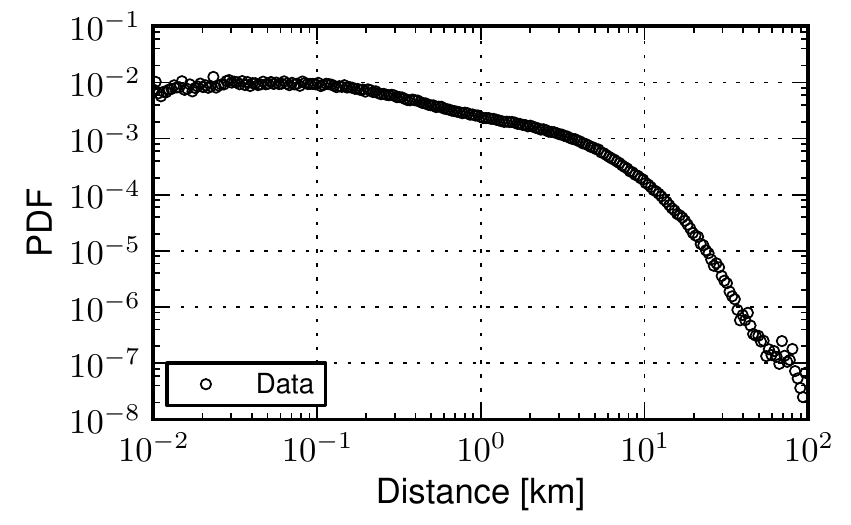}
\label{intracity}
}
\caption{\textbf{Global and urban movements.} (a) The probability density function (PDF) of human displacements as seen through 35 million location broadcasts (checkins) across the planet. The power-law fit features an exponent $\beta = 1.50$ and confirms previous works on human mobility data. The spatial granularity offered by GPS data allows for the inspection of human movements at very small distances, whereas the global reach of Foursquare reveals the full tail of the planetary distribution of human movements. (b) The probability density function (PDF) of human displacements in cities (intracity). For two successive location broadcasts (checkins) a sample is included if the locations involved in the transition belong to the same city. Approximately 10 million of those transitions have been measured. The poor power-law fit of the data ($\beta=4.67$, $\Delta r_{0}=18.42$)  suggests that the distribution of intracity displacements can not be fully described by a power law. Short  transitions which correspond to a large portion of the movements distribution are not captured by such process.}
\end{figure*}

\label{sec:intro}
Since the seminal works of Ravenstein \cite{ravenstein}, the movement
of people in space has been an active subject of research in the social and
geographical sciences. It has been shown in almost every quantitative study and
described in a broad range of models that a close relationship exists between
mobility and distance. People do not move randomly in space, as we know from our
daily lives. Human movements exhibit instead high levels of regularity and tend
to be hindered by geographical distance. The origin of this dependence of
mobility on distance, and the formulation of quantitative laws explaining human
mobility remains, however, an open question, the answer of which would lead to
many applications, e.g. improve engineered systems such as cloud
computing and location-based recommendations \cite{zheng2009,zheng,quercia,scellato},
enhance research in social networks \cite{onnela,crandall,scellato2011,leskovec} and yield
insight into a variety of important societal issues, such as urban planning and
epidemiology \cite{nicholson,hufnagel,colizza}.

In classical studies, two related but diverging viewpoints have emerged. The
first camp argues that mobility is directly deterred by the costs (in time and
        energy) associated to physical distance. Inspired by Newton's law of
gravity, the flow of individuals is predicted to decrease with the physical
distance between two locations, typically as a power-law of distance
\cite{gr1,gr3,gr5}. These so-called ``gravity-models" have a long tradition in
quantitative geography and urban planning and have been used to model a wide
variety of social systems, e.g. human migration \cite{gravity1b}, inter-city communication \cite{krings} and traffic
flows \cite{gravity2}. The second camp argues instead that there is no direct
relation between mobility and distance, and that distance is a surrogate for the
effect of {\em intervening opportunities} \cite{stouffer}. The migration from
origin to destination is assumed to depend on the number of opportunities closer
than this destination. A person thus tends to search for destinations where to
satisfy the needs giving rise to its journey, and the absolute value of their
distance is irrelevant. Only their ranking matters. Displacements are thus
driven by the spatial distribution of places of interest, and thus by the
response to opportunities rather than by transport impedance as in gravity
models. The first camp appears to have been favoured by practitioners on the
grounds of computational ease \cite{easa}, despite the fact that several
statistical studies have shown that the concept of intervening opportunities is
better at explaining a broad range of mobility data
\cite{miller,haynes,wadicky,Freymeyer,cheung}.

This long-standing debate is of particular interest in view of the recent
revival of empirical research on human mobility. Contrary to traditional works,
        where researchers have relied on surveys, small-scale observations or
        aggregate data, recent research has taken advantage of the advent of
        pervasive technologies in order to uncover trajectories of millions of
        individuals with unprecedented resolution and to search for universal
        mobility patterns, such to feed quantitative modelling. Interestingly,
        those works have all focused on the probabilistic nature of movements in
        terms of physical distance. As for gravity models, this viewpoint finds
        its roots in Physics, in the theory of anomalous diffusion. It tends to
        concentrate on the distributions of displacements as a function of
        geographic distance. Recent studies suggest the existence of a universal
        power-law distribution  $P(\Delta r) \sim \Delta r^{-\beta}$, observed
        for instance in cell tower data of humans carrying mobile phones $\beta
        = 1.75$ \cite{marta} or in the movements of ``Where is George" dollar
        bills $\beta = 1.59$ \cite{Brockmann2006}. This universality is,
        however, in contradiction with observations that displacements strongly
        depend on where they take place. For instance, a study of hundreds of
        thousands of cell phones in Los Angeles and New York demonstrate
        different characteristic trip lengths in the two cities \cite{isaacman}.
        This observation suggests either the absence of universal patterns in
        human mobility or the fact that physical distance is not a proper
        variable to express it. 

In this work, we address this problem by focusing on human mobility patterns in
a large number of cities across the world. More precisely, we aim at answering
the following question: ``Do people move in a substantially different way in
different cities or, rather, do movements exhibit universal traits across disparate
urban centers?".  To do so, we take advantage of the advent of mobile
location-based social services accessed via GPS-enabled smartphones, for which
fine granularity data about human movements is becoming available. Moreover, the
worldwide adoption of these tools implies that the scale of the datasets is
planetary. Exploiting data collected from public \textit{check-ins} made by
users of the most popular location-based social network,
      Foursquare~\cite{foursquare}, we study the movements of 925,030
      users around the globe over a period of about six months, and study the
      movements across 5 million places in 34 metropolitan cities that
      span four continents and eleven countries. 

After discussing how at larger distances we are able to reproduce previous
results of \cite{marta} and \cite{Brockmann2006}, we also offer new insights on
some of the important questions about human urban mobility across a variety of
cities.  We first confirm that mobility, when measured as a function of
distance, does not exhibit universal patterns. The striking element of our
analysis is that we observe a universal behavior in all cities when measured
with the right variable. We discover that the probability of transiting from one
place to another is inversely proportional to a power of their {\em rank}, that
is, the number of intervening opportunities between them. This universality is
remarkable as it is observed despite cultural, organizational and national
differences. This finding comes into agreement with the social networking
parallel which suggests that the probability of a friendship between two
individuals is inversely proportional to the number of friends between
them~\cite{liben}, and depends only indirectly on physical distance. More importantly,
    our analysis is in favour of the concept of intervening opportunities rather
    than gravity models, thus suggesting that trip making is not explicitly
    dependent on physical distance but on the accessibility of objectives
    satisfying the objective of the trip. Individuals thus differ from random
    walkers in exploring physical space because of the motives driving their
    mobility.

Our findings are confirmed with a series of simulations verifying the hypothesis
that the place density is the driving force of urban movement.  By using only
information about the distribution of places of a city as input and by coupling
this with a rank-based mobility preference we are able to reproduce the actual
distribution of movements observed in real data. These results open new
directions for future research and may positively impact many practical systems
and application that are centered on mobile location-based services.

\section{Results}
\subsection{Urban Movements and Power-laws}
\begin{figure}[t]
\centerline{\includegraphics[scale=1.0]{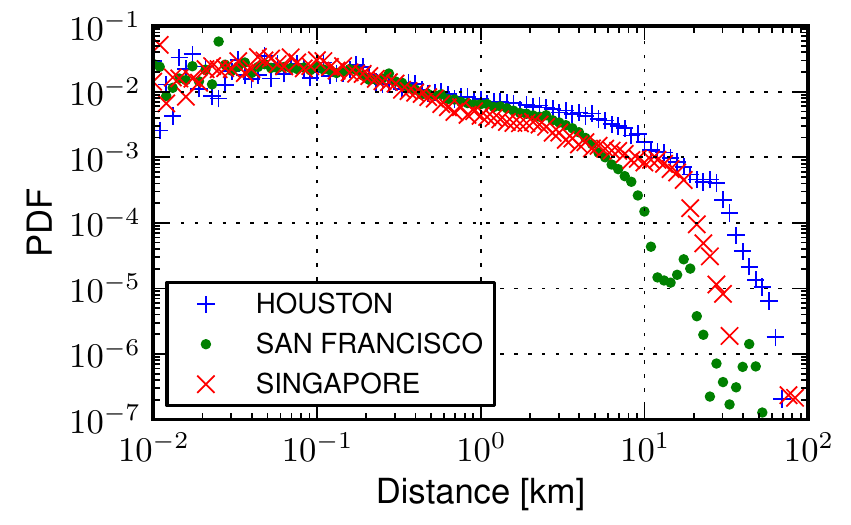}}
\caption{The probability density function (PDF) of human displacements in three
    cities: Houston, San Francisco and Singapore (for 47, 112 and 79 thousand
            transitions, respectively). Common trends are observed, e.g., the
        probability of a jump steadily decreases after the distance threshold of
        100 meters, but the shapes of the distributions vary from city to city,
              suggesting either that human movements do not exhibit universal
                  patterns across cities or that distance is not the appropriate
                  variable to model them.}
\label{manycitiesmovements}
\end{figure}

\label{sec:urban}
We draw our analysis upon a dataset collected from the largest Location-based
Social Network, Foursquare~\cite{foursquare}.  The dataset features 35,289,629
movements of 925,030 users across 4,960,496 places collected during six months
in 2010. 
By movements we express the indication of presence at a place that a user gives through the system (in the language of Location-based Social
    Networks, a location broadcast is referred as a \textit{checkin}). For each
    place we have exact GPS coordinates.

In order to confirm the large scale results reported in
\cite{marta,Brockmann2006}, we have computed the distribution of human
displacements in our dataset (Figure \ref{alltrans}): we observe that the
distribution is well approximated by a power law with exponent $\beta = 1.50$ ($p-value = 0.494$).
This is almost identical to the value of the
exponent calculated for the dollar bills movement ($\beta = 1.59$)  \cite{Brockmann2006} and very proximate to the $1.75$ estimated from cellphones calls
analysis of human mobility \cite{marta}. 
With respect to these datasets, we note that the Foursquare dataset is planetary, as it contains movements at distances up to 20,000
kilometres (we measure all distances
using the great-circle distance between points on the planet). On 
the other extreme, small distances of the order of tens of meters can also be
approximated thanks to the fine granularity of GPS technology employed by mobile phones running these geographic social network applications. Indeed, we find that the
probability of moving up to 100 meters is uniform, a trend that has also been shown in~\cite{Brockmann2006} for 
a distance threshold $\Delta r_{min}$. Each
transition in the dataset happens between two well defined venues, with data
specifying the city they belong to. We exploit this
 information  to define when a transition is
urban, that is, when both start and end points are located within the same city. Figure \ref{intracity} depicts the probability density function of the
about 10 million displacements within cities across the globe. 
We note that a power-law fit does not accurately capture the distribution. First
of all, a large fraction of the distribution exhibits an initial flat trend; 
then, only for values larger than 10 km the tail of distribution decays, albeit with a very
large exponent which does not suggest a power-law tail. Overall, power-laws tend
to be captured across many orders of magnitude, whereas this is not true in the case
of urban movements. The estimated parameter values are $\Delta r_{0} = 18.42$ and exponent $\beta=4.67$ ($p-value = 1.0$). 

\begin{figure*}[t]
\subfigure
{
\includegraphics[scale =1] {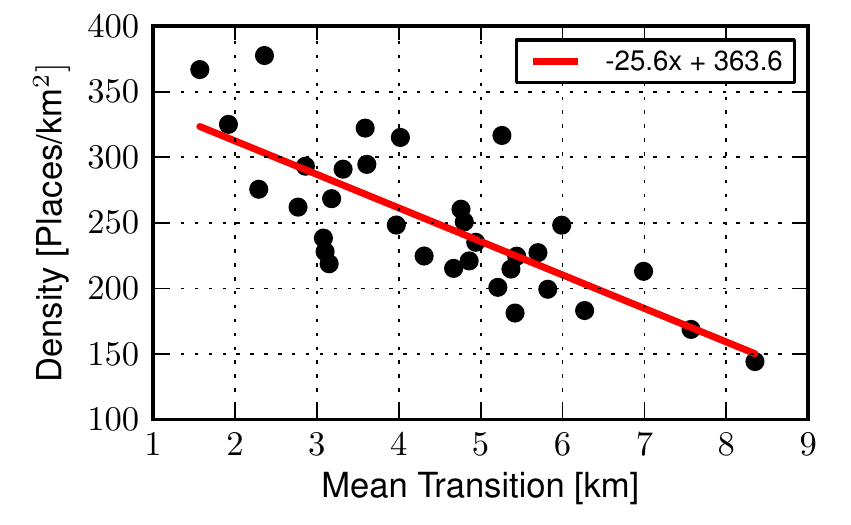}
\label{density}
 }
\subfigure
{
\includegraphics[scale =1] {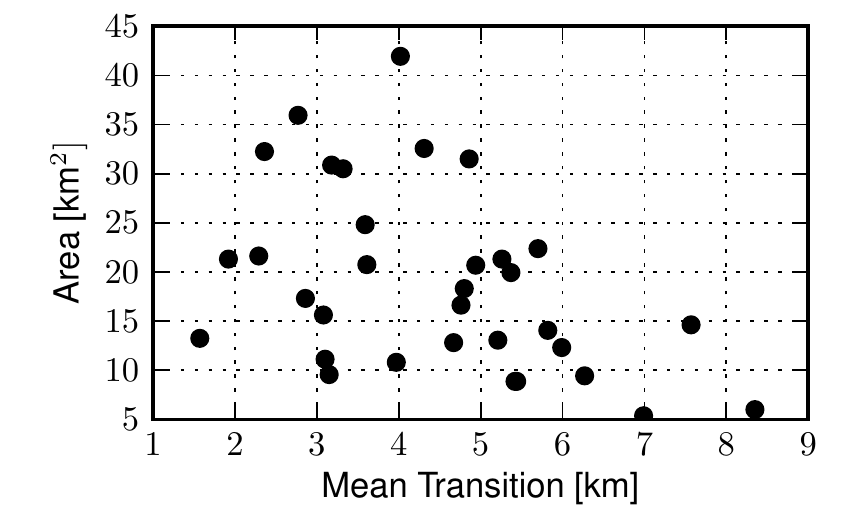}
\label{area}
}
\caption{\textbf{Geographical characteristics and mean urban movements.} (a) Scatter plot of the density of a city, defined as the number of places per square kilometer, versus its mean human transition in kilometers. Each datapoint corresponds to a city, while the red line is a fit  that highlights the relationship of the two variables ($R^2=0.59$). A longer mean transition corresponds to the expectation of a sparser urban environment, indicating that the number of available places per area unit could have an impact on human urban travel. (b) Scatter plot of the area of a city, measured in square kilometers, versus its mean human transition in kilometers. Unlike place density, the area of a city does not seem strongly related to the mean length of its transitions ($R^2=0.19$). To measure the area of a city we have segmented the spatial plane around its geographic midpoint in squares of size $250 \times 250$ $m^{2}$. The area of a city has been defined as the sum area of all squares that feature at least five places.}
\end{figure*}

%
%
%

\subsection{Movements across cities.}
\label{sec:movements}
Since the distribution of urban human movements cannot be approximated with a power
law distribution nor with a physically relevent functional relation, how can we represent displacements of people in a city more appropriately? 
We start by
comparing human movements across different cities. In
Figure \ref{manycitiesmovements}, we plot the distribution of human
displacements for a number of cities. The shapes of the distributions, albeit different, exhibit similarities suggesting
 the existence of a common underlying process
that seems to characterize human movements in urban environments. There is an almost
uniform probability of traveling in the first 100 meters, that is followed by a
decreasing trend between 100 meters and a distance threshold $\delta_{m} \in
[5,30]$ km, where we detect an abrupt cutoff in the probability of observing a
human transition.  The threshold $\delta_{m}$ could be due to the reach of the
\textit{borders} of a city, where maximum distances emerge. 

While the distributions exhibit similar trends in different cities, scales and functional relation may differ, thus suggesting that human mobility vary from city to city. For example, while comparing
Houston and San Francisco (see Figure \ref{manycitiesmovements}), different thresholds $\delta_{m}$ are observed. Moreover, the probability densities
can vary across distance ranges. For instance, it is more probable to have a transition in the
range 300 meters and 5 kilometers in San Francisco than in Singapore, but the opposite is true  beyond 5 kilometers. This difference could be attributed to many
      potential factors, ranging from geographic ones such as area
      size, density of a city, to differences in infrastructures such as
      transportation and services or even socio-cultural variations across
      cities. In the following paragraphs we present a formal analysis that allow
to dissect these heterogeneities. 

\subsection{The importance of place density.}
\label{sec:density}
Inspired by Stouffer's theory of intervening opportunities~\cite{stouffer} which suggests
that \textit{the number of persons traveling a given distance is directly proportional to the number of opportunities at that distance and inversely proportional to the number of intervening opportunities}, 
we explore to what extend the density of
places in a city is related to the human displacements within it. As a
first step, we plot the place density of a city, as computed with our checkin
data, against the average distance of
displacements observed in a number of cities. In Figure \ref{density} one observes that the average distance
of human movements is {\em inversely proportional} to the city's density. Hence, in a
very dense metropolis, like New York, there is a higher expectation of shorter
movements. We have measured a coefficient of determination $R^2=0.59$. Intuitively, this correlation suggests that while distance is a cost factor
taken into account by humans, the range of available places at a given distance
is also important. This availability of places may relate to the availability of
resources while performing daily activities and movements: if no
super markets are around, longer movements might be more probable in order to find supplies.
As a next step, we explore whether the geographic area size covered by a city
affects human mobility by plotting 
the average transition in a city versus its area size (see Fig. \ref{area}). Our data indicates 
 no apparent linear relationship, with a low correlation $R^2=0.19$, thus indicating that density is a more informative
measure.

\begin{figure*}[t]
\subfigure
{
\includegraphics[scale =1] {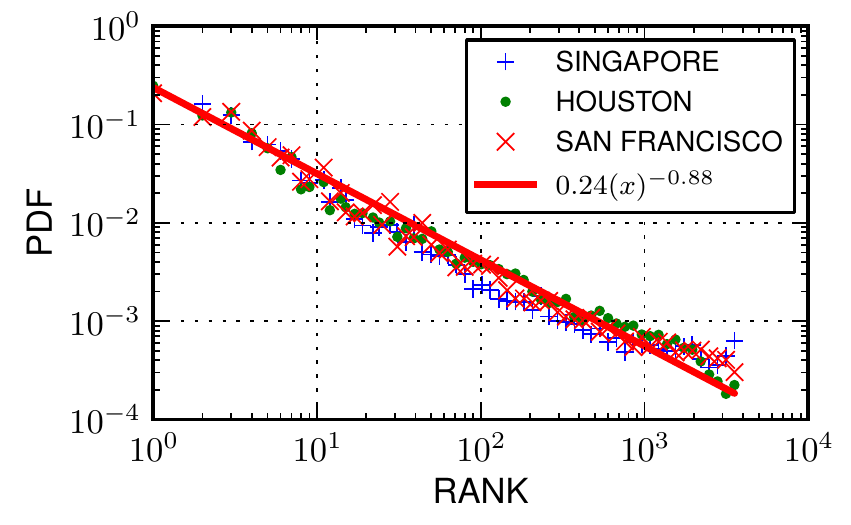}
\label{threecitiesranks}
 }
\subfigure
{
\includegraphics[scale =1] {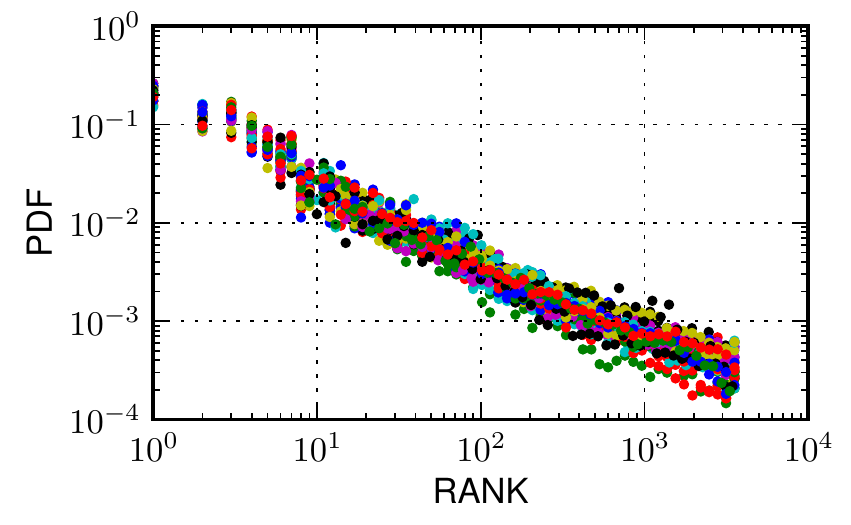}
\label{manycitiesranks}
}
\caption{\textbf{Rank distributions across urban environments.} (a) Probability density function (PDF) of rank values for three cities
    (Houston, Singapore and San Francisco). Our methodology to measure the rank
        distribution is the following: for each transition between two places
        $u$ and $v$, we measure $rank_{u}(v)$ defined as the number of places that
        are geographically closer to $u$ than $v$. We observe that the
        distributions of the three cities collapse to a single line, which suggests
        that 
        universal laws can be formulated in terms of the rank variable. The observation confirms the hypothesis that
        human movements are driven by the density of the geographic environment
        rather than the exact distance cost of our travels. A least squares fit
        (red line) underlines the decreasing trend of the probability of a jump
        as the rank of a places increases. (b) Superimposition of the probability density functions (PDF) of rank
		    values the thirty-four cities analyzed in the Foursquare dataset. A decreasing trend for
		        the probability of a jump at a place as its rank value increases is
		        common. The trend remains stable despite the large number of plotted
		        cities and their potential differences with respect to a number of variables such
		        us number of places, number of displacements, area size, density or
		        other cultural, national or organizational ones.}
\end{figure*}

To shed further light on the hypothesis that density is a decisive factor in human mobility, for every movement between
a pair of places in a city we sample the rank value of it. The rank for each
transition between two places $u$ and $v$ is the number of places $w$ that are 
closer in terms of distance to $u$ than $v$ is. Formally: $rank_{u}(v) = |\{
w:d(u,w) <d(u,v)\}|.$
The rank between two places has the important property to be invariant in scaled versions of a city, where the relative positions of the places is preserved but the absolute distances dilated.
In Figure \ref{threecitiesranks} we plot for the three cities the rank values observed for each
displacement. The fit of the rank densities on a log-log plot, shows that the rank distribution follows linear trend
similar to that of a power-law distribution. This observation suggests that the probability of moving to a place
decays when the number of places nearer than a potential destination increases. 
Moreover, the ranks of all cities collapse on the same line despite the variations
in the probability densities of human displacements. 
We have fit the rank distribution for the thirty-four cities under investigation and have measured an
exponent $\alpha=0.84 \pm 0.07$. This is indicative of a universal pattern across cities where
density of settlements is the driving factor of human mobility. We superimpose the distribution
of ranks for all cities in Figure \ref{manycitiesranks}.

Interestingly enough, a parallel of this finding can be drawn with the results
in~\cite{liben}, where it is found that the probability of observing a user's friend
at a certain distance in a geographic social network is inversely proportional to the number of people
geographically closer to the user.

\subsection{Modelling urban mobility.}
\label{sec:rankmodel}
The universal mobility behaviour emerging across cities paves the way
to a new model  of movement in urban environments.
Given a set of places ${\cal U}$ in a city, the probability of moving
from place $u\in {\cal U}$ to a place $v \in {\cal U}$ is formally defined as

\begin{equation*}
	Pr[u \rightarrow v] \propto \frac{1}{ rank_{u}(v)^{a} }
\end{equation*}
where 
\begin{equation*}
	rank_{u}(v) = |\{ w:d(u,w) <d(u,v)\}|.
\end{equation*}

\begin{figure*}[t]
\includegraphics[scale=1.4]{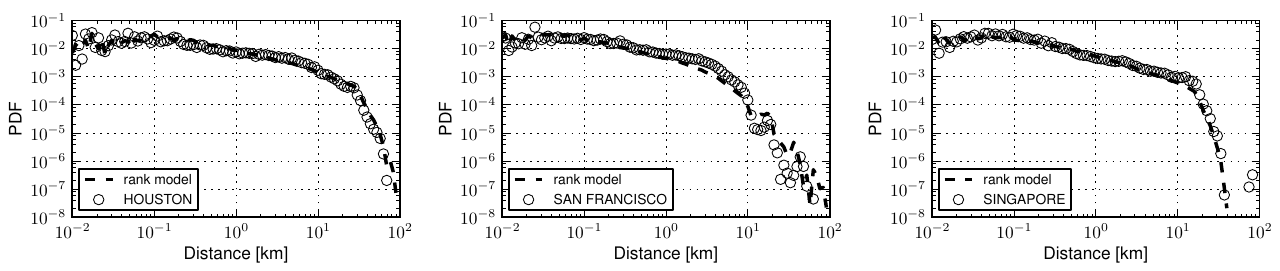}
\caption{Probability Density Functions (PDF) of human movements and
    corresponding fits with the rank-based model in three cities (Houston, San
            Francisco and Singapore). In the rank-based model the probability of
        transiting from a place $u$ to a place $v$ in a city, only depends on
        the rank value of $v$ with respect to $u$. The places of a city employed
        for the simulation experiments where those observed in the Foursquare
            dataset, hence while the rank-based model is the same for all cities
                the underlying spatial distribution of places may vary. Excellent fits are observed for all
                cities analyzed. It
                is interesting to note that the model is able to reproduce even
                minor anomalies, such as the case of San Francisco where we have
                'jumps' in the probability of a movement at 20 and 40
                kilometers.}
\label{manycitiesmovementsfits}
\end{figure*}

We run agent based simulation experiments (see detailed description in \cite{supporting}) where agents transit from one
place to another according to the probabilities defined by the model
above. Averaging the output of the probability of movements by considering all possible places of a city as potential starting points
for our agents, we present the human displacements resulting from the
model in Figure~\ref{manycitiesmovementsfits}: 
as shown, despite the simplicity of the model, this is able to capture with very
high accuracy the real human displacements in a city.  
Our model does not take into account other
parameters such as individual heterogeneity patterns~\cite{marta} or temporal ones~\cite{Brockmann2006} that have been studied in the past in the context of human mobility and yet it offers very accurate matching of the human traces of our dataset. A common parameter $\alpha=0.84$ (empirical average) has been set for the simulations of all cities. We have observed movement fits to deteriorate as we move away from values measured empirically. In general, large values overestimate trips to nearby places and inversely very low $\alpha$ values blur the effect of distance in human movements. 

\begin{figure*}[t]
\includegraphics[scale=1.1]{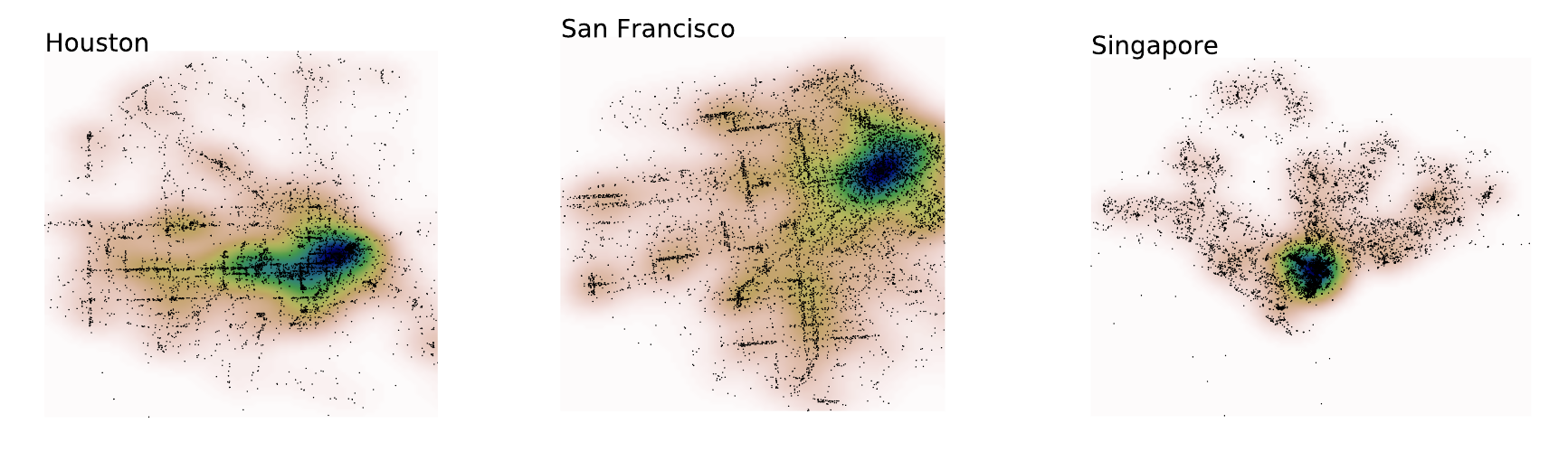}
\caption{Gaussian kernel density estimation (KDE) applied on the spatial distribution of places in three cities (Houston, San Francisco and Singapore). Each dark point corresponds to a venue observed in the Foursquare dataset encoded in terms of longitude and latitude values. The output of the KDE is visualised with a thermal map. A principal core of high density is observed in the three cities, but point-wise density and spatial distribution patterns may differ. The rank-based model can cope with those heterogeneities as it accounts for the relative density for a given pair of places $u$ and $v$.}
\label{manycitiesMaps}
\end{figure*}

\begin{figure}[t]
\centerline{\includegraphics[scale=1.0]{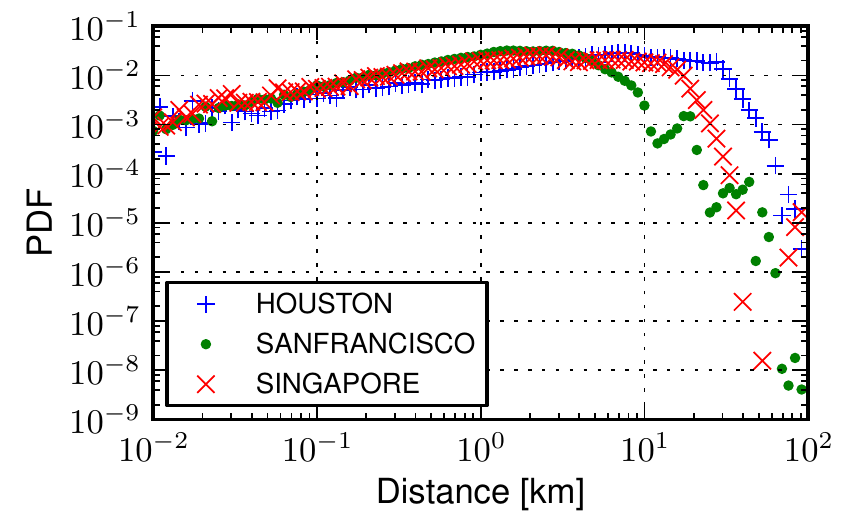}}
\caption{Probability density function (PDF) of observing two randomly selected places at a distance $\Delta r$  in a city. We have enumerated 11808, 15970, 15617 unique venues for Houston, San Francisco and Singapore respectively. The probability is increasing with $\Delta r$, as expected in two dimensions before falling due to finite size effect. It is interesting to note that the probability for two randomly selected places to be the origin and destination of a jump monotonically decreases with distance (see SI).}
\label{placecities}
\end{figure}

Moreover, our analysis provides empirical evidence that while human displacements across
cities may differ, these variations are mainly due to the spatial distribution of
places in a city instead of other potential factors such as social-cultural or
cognitive ones. Indeed, the agent based simulations are run with the same rules and parameters in each city, except for the set of places
${\cal U}$ that is taken from the empirical dataset. The variation across the spatial organization of cities is illustrated in Fig. \ref{manycitiesMaps}, where we plot thermal maps of the density of places within cities and in Fig. \ref{placecities}, where we plot the probability density function that two random places are at a distance $\Delta r$. Our analysis highlights the impact of geography, as expressed through the spatial distribution 
of places, on human movements, and confirms at a large-scale the seminal analysis of Stouffer~\cite{stouffer} who studied how the spatial distribution of places in the city of Cleveland affected the migration movements of families. Our analysis does not indicate that distance does not play a deterring role, but that it is not sufficient to express human mobility through universal laws. A proper modeling should account for place distribution, as rank does or possibly by complementing distance with information about place locations, as in constrained gravity models \cite{gr3}.
Plots for all thirty four cities that we have evaluated can be found in SI~\cite{supporting}.

\section{Discussion}
The empirical data on human movements provided by Foursquare and other
location-based services allows for unprecedented analysis both in terms of scale
and the information we have about the details of human movements. The former
means that mobility patterns in different parts of the world can be analyzed
and compared surpassing cultural, national or other organizational borders. The
latter is achieved through better location specification technologies such as
GPS-enabled smartphones, but also with novel online services that allow users to
layout content on the geographical plane such as the existence of places and
semantic information about those. As those technologies advance our
understanding on human behavior can only become deeper. 

In this article, we have focused on human mobility in a large number of metropolitan cities around
the world to perform an empirical validation of  past
theories on the driving factors of human movements. As we have shown,
 Stouffer's~\cite{stouffer} theory of intervening opportunities appears
to be a plausible explanation to the observed mobility patterns. The
theory suggests that the distance covered by humans is determined by
the number of opportunities (i.e., places) within that distance, and not by the distance itself. 
This behaviour is confirmed in our data where we observed that physical distance does not allow for the 
formulation of universal rules for human mobility, whereas a universal pattern emerges across all
cities when movements are analyzed through their respective rank values: the probability of a transition to a destination place is inversely
proportional to the relative rank of it, raised to a power $\alpha$, with
respect to a starting geographical point. Moreover, $\alpha$ presents minor
variations from city to city. 

We believe that our approach opens avenues of quantitative exploration of human mobility, with several applications in urban planning and ICT.
The identification of rank as an appropriate variable for the deterrence of human mobility is in itself an important observation, as it is expected to lead to more reliable measurements in systems where the density of opportunities is not uniform, e.g. in a majority of real-world systems. The realization of universal properties in cities around the globe also goes along the line of recent research~\cite{bettencourt2007,bettencourt2010} on urban dynamics and organization, where cities have been shown to be scaled versions of each other, despite their cultural and historical differences. Contrary to previous observations where size is the major determinant of many socio-economical characteristics, however, density and spatial distribution are the important factors for mobility. Moreover, the richness of the dataset naturally opens up new research directions, such as the identification of the needs and motives driving human movements, and the calibration of the contact rate, e.g. density- vs frequency-dependent, in epidemiological models \cite{smith}. 

\section{Methods}
\label{sec:methods}
The mobility dataset used in this work is comprised from \textit{checkins} made by Foursquare users 
and become publicly available through Twitter's Streaming API. The collection process lasted from the 27th of May
2010 until the 3rd of November of the same year. During this period we have observed 35,289,629 \textit{checkins}
from 925,030 unique users over 4,960,496 venues. 
In addition, \textit{locality} information together with exact GPS geo-coordinates for each venue has become available through the Foursquare website allowing us
to associate a given venue with a city. By considering only consecutive \textit{checkins} that take place within the same city
we have extracted almost 10 million \textit{intracity} movements analysed in Figure \ref{intracity}. Detailed statistics
including the number of checkins and venues in each city can be found in~\cite{supporting}.

We have employed the methods detailed in \cite{clauset} to apply goodness-of-fit tests on the Probability Density Functions of global and urban transitions observed in Figures \ref{alltrans} and \ref{intracity}. In particular, we have measured the corresponding $p-values$ using the Kolmogorov-Smirnov test by generating 1000 synthetic distributions, while the Maximum-Likelihood Estimation technique has been used to estimate the parameters of the power-laws. Exceptionally, we have resorted to a \textit{least squares} based optimization to measure the exponent $\alpha$ of the rank values shown in Figure \ref{threecitiesranks}. While power-laws are not well defined for exponents smaller than $1$, we are confident of the values estimated due to the excellent movement fits produced during our simulations. 


\begin{acknowledgments}
RL thanks M. Gonzales for mentioning him the original work of Stouffer. This research was supported in part by the National Science Foundation under Grant No. NSF PHY05-51164.
\end{acknowledgments}





\end{document}